\documentclass[a4paper,11pt]{article}
\pdfoutput=1 

\usepackage{jcappub} 

\usepackage[T1]{fontenc} 

\title{\boldmath Dark Matter directional detection: comparison of the track direction determination}

 \author{C.~Couturier}
 \author{J. P.~Zopounidis}
 \author{N.~Sauzet}
 \author{F.~Naraghi}
 \author{and D.~Santos}
\affiliation{LPSC, CNRS/IN2P3, UJF, INP, Avenue des Martyrs, 38000, Grenoble, France}

\emailAdd{ccouturi@lpsc.in2p3.fr}
\emailAdd{zopounidis@lpsc.in2p3.fr}
\emailAdd{sauzet@lpsc.in2p3.fr}
\emailAdd{naraghi@lpsc.in2p3.fr}
\emailAdd{santos@lpsc.in2p3.fr}

\abstract{
	 Several directional techniques have been proposed for a directional detection of Dark matter, among others anisotropic crystal detectors, nuclear emulsion plates, and low-pressure gaseous TPCs. The key point is to get access to the initial direction of the nucleus recoiling due to the elastic scattering by a WIMP.
	 In this article, we aim at estimating, for each method, how the information of the recoil track initial direction is preserved in different detector materials.
	 We use the SRIM simulation code to emulate the motion of the first recoiling nucleus in each material.
	 We propose the use of a new observable, \textit{D}, to quantify the preservation of the initial direction of the recoiling nucleus in the detector. We show that in an emulsion mix and an anisotropic crystal, the initial direction is lost very early, while in a typical TPC gas mix, the direction is well preserved.
	}


\begin{document}
\maketitle
\flushbottom

\section{Introduction}

Weakly interacting massive particles are among the most studied candidates for Dark Matter (DM). The direct search for DM is performed by looking for elastic collisions of these WIMPs with target nuclei in a detector~\cite{goodman_1985}. For a 1\,GeV/c$^2$ to 1000\,GeV/c$^2$ DM particle, the kinetic energy of the recoiling nuclei would be in the range of few keV to few hundreds of keV depending on the mass of the target nuclei.
The search for such rare events, at low energies, requires a highly efficient discrimination of the background with respect to signal composed of nuclear recoils. 

Non-directional detection is mainly limited -- after an ideal electron-recoil discrimination -- by two non-reducible background components: the neutrons and the neutrinos. 
Neutron background comes from natural radioactivity of the detector itself and the surrounding environment ($^{238}$U and $^{232}$Th) and from cosmic muons interacting with the rock and the detector. This background is present in all detectors, no matter how large the fiducial volume is. A veto can help to partly discriminate this background ; however the neutrons produced in the shielding or inside the detector are hard to discriminate.
Neutrinos may come from the Sun, the atmosphere (interaction with cosmic rays) and from the DSNB (Diffuse Supernova Neutrino Background). The coherent scattering of these neutrinos on the target nuclei produces recoils in the energy region of interest (\mbox{E$_R$ $\lesssim$ 100\,keV}). When scaling the experiment, the neutrino background becomes non negligible.
This ultimate neutrino background, called ``neutrino floor'', cannot be discriminated with standard non-directional methods, as no veto or shielding can be set for neutrinos and thus enforces a lower limit to the cross section attainable with non-directional detectors~\cite{billard_2014}.

On the other hand, directional detection proposes to use the  anisotropy in the recoil angular distribution (in the galactic coordinates) originated from the  motion of the solar system around the galactic center within the DM halo~\cite{spergel_1988}. The mostly isotropic background can thus be distinguished from the (anisotropic) expected DM signal.
In case of a claimed detection by a non-directional detector, a directional detection would be an unambiguous proof of the DM origin of the claimed signal~\cite{billard_2012}.

The crucial point of the directional detection is to get access to the initial direction of the recoil nucleus. 
Several strategies have been proposed for a directional detection. 
We will focus on three strategies and present them: anisotropic crystals, nuclear emulsions and low pressure gaseous TPCs.
The goal here is to provide a method to assess how the information of the initial direction of a nuclear recoil is retained in a given material. 
The direction actually measured by a detector is mainly defined by the WIMP-ion recoil angular distribution. Instrumental effects might dilute the initial direction information: readout \cite{battat_readout_2016}, diffusion effects. 
We will avoid the WIMP-ion kinematic effects, and will focus on the motion of the first recoiling nucleus in the detector material, and consider it has a fixed energy and a fixed direction. We use SRIM simulations~\cite{ziegler_2008,ziegler_2010} to emulate the motion of such nuclear recoil in each detector material; the SRIM outputs will help us define a figure of merit evaluating the preservation of the direction of a nuclear recoil in each material.

\section{Principle of detection}\label{sec:pod}

\subsection{Crystal
	detectors}\label{sub:pod-crystal}
Several groups have proposed anisotropic crystal detectors for a ``directional detection'', using \textit{e.g.} stilbene crystals \citep{sekiya_measurements_2003} or ZnWO$_4$ \citep{cappella_e_2013}. 
The response (pulse shape, light output) of particles recoiling on the anisotropic crystal depends on the direction of the particles with respect to the crystal axes. 
This crystal detector is not able to measure individual tracks, but instead relies on the statistical analysis of the number of events in different crystals oriented along the three different crystallographic planes.

\subsection{Emulsion detectors}\label{sub:pod-emulsion}
Emulsion plates have been used widely in particle physics to measure the tracks of charged particles. 
D'Ambrosio \textit{et al.} \cite{dambrosio_2014} report ongoing R{\&}D to show the feasibility of an efficient directional detection of DM-induced nuclear recoils using emulsion layers, mounted on an equatorial telescope to keep them in the same direction towards the Cygnus constellation (layers parallel to the expected WIMP ``wind''). Their emulsion is made of AgBr crystal in a gel. The crystal grain size, as well as the readout system, determine the precision of the track measurement. Films using grain of a size of about 40\,nm have been developed (Nano Imaging Trackers). 
Recent developments of fully automated readout systems have made it possible to fast scan the emulsion layers in optical, to make a first selection of the tracks with a $\sim 100$\,nm precision on the track length measurement. The remaining tracks are then analyzed with a X-ray readout with a claimed precision of $50$\,nm.

\subsection{Time projection chambers (gaseous
	detectors)}\label{sub:pod-tpc}
Several TPC directional prototypes are operating underground: DRIFT in Boulby \citep{daw_e_2012}, DMTPC at WIPP \citep{battat_j_2013},
Newage at Kamioka laboratory \citep{miuchi_e_2012} and MIMAC in Modane \citep{santos_j_2013}.
They use gaseous mixtures 
at low pressure to get long enough  recoil tracks at low energies. 
Two types of gas mixtures can be distinguished: mixtures in which we collect electrons, for instance CF$_4$ based mixtures, with fast electron drift velocities but a high transverse diffusion. Electronegative gas such as CS$_2$ allow for negative ions drifts with less dispersion, thus possible over longer distances.

A pixelated readout plane allows the measurement of the 2D-projection of the ionization signal conveyed by primary electrons/ions (DMTPC).
The reconstruction of the track in 3D is obtained with a time-sampling of the 2D-projection (DRIFT-II, Newage, MIMAC).

\section{Simulations}\label{sec:comparison}

Our aim here is to compare only one aspect of a directional detection: how well the directional information is kept in different materials. To do this comparison, independently of the detector used to study this phenomenon, we chose a common simulation tool, SRIM~\cite{ziegler_2008,ziegler_2010}. 
We follow these steps:
\begin{enumerate}
	\item we simulate nuclear recoils with a fixed kinetic energy, and a fixed angle ($\theta = 0$), to feed into the SRIM simulation. SRIM then generates the nuclear recoil track.
	\item using track segments from SRIM, we define an observable, \textit{D}, that describes how well the directional information is conserved.
\end{enumerate}

\subsection{Models for the SRIM simulations}\label{sub:models}

\begin{itemize}
	\item
	Crystal
	scintillators: We consider oxygen recoils in a ZnWO\(_4\) detector; the density of this crystal is 7.87 g/cm\(^3\). 
	Channeling has been reported as a weak effect \cite{lindhard_1965,bozorgnia_2010}, so this effect has not been taken into account in these simulations.
	\item
	Emulsion: We consider an emulsion mix composed of the following nuclei (mass fraction in \%): 
	H (1.63), C (10.12), N (2.68), O (7.40), S (0.03), Br (32.20), Ag (44.07), I (1.87). 
	The resulting mix has a density of 3.2 g/cm\(^3\). 
	In the proposed emulsion mix, Ag and Br could be interesting targets in the search for higher mass WIMPs (100-1000 GeV) ; however, their very short range would make them undetectable by the dedicated readout systems. On the other hand, C, N are expected to have larger ranges, additionally to be more suited to lower mass WIMPs search (10 GeV). We consider the most favorable case in the simulation, and study the carbon recoils in the emulsion mix.
	No threshold has been applied in the analysis of the tracks.
	\item
	TPCs: We consider a volume of gas composed of 70\% CF\(_4\) + 28\% CHF\(_3\) + 2\% C\(_4\)H\(_{10}\) ; 
	this gas mix is in use in the MIMAC prototype at Modane. 
	The pressure is fixed at \mbox{50 mbar}. This leads to a density of 1.72 10\(^{-4}\) g/cm\(^3\).
\end{itemize}

\subsection{Simulation outputs}
\begin{figure}[htbp]
	
	\centering
	\includegraphics[width=0.328\linewidth]{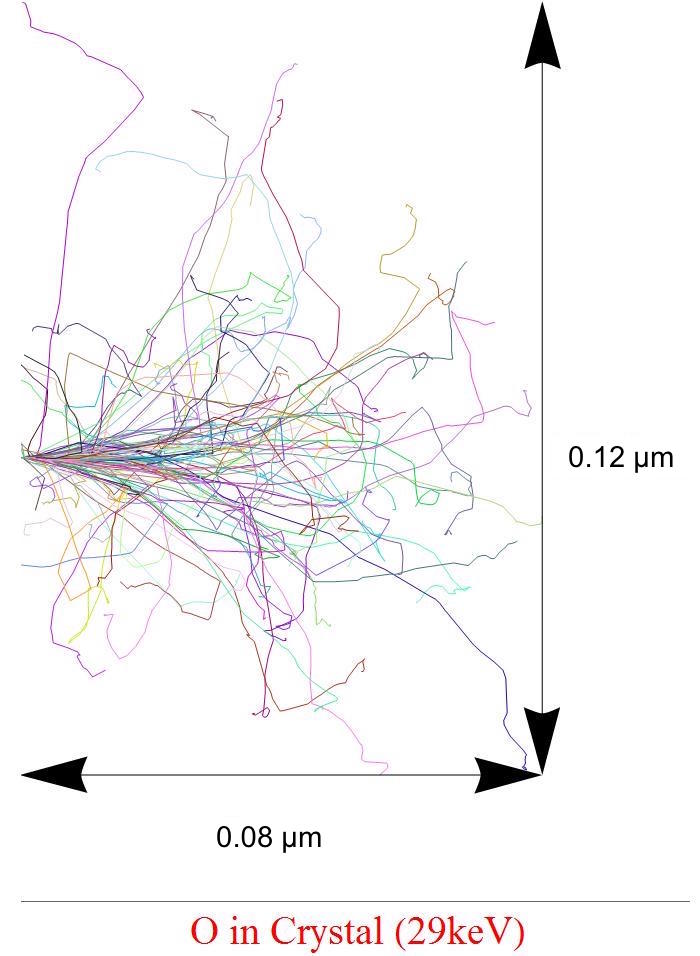}
		\includegraphics[width=0.328\linewidth]{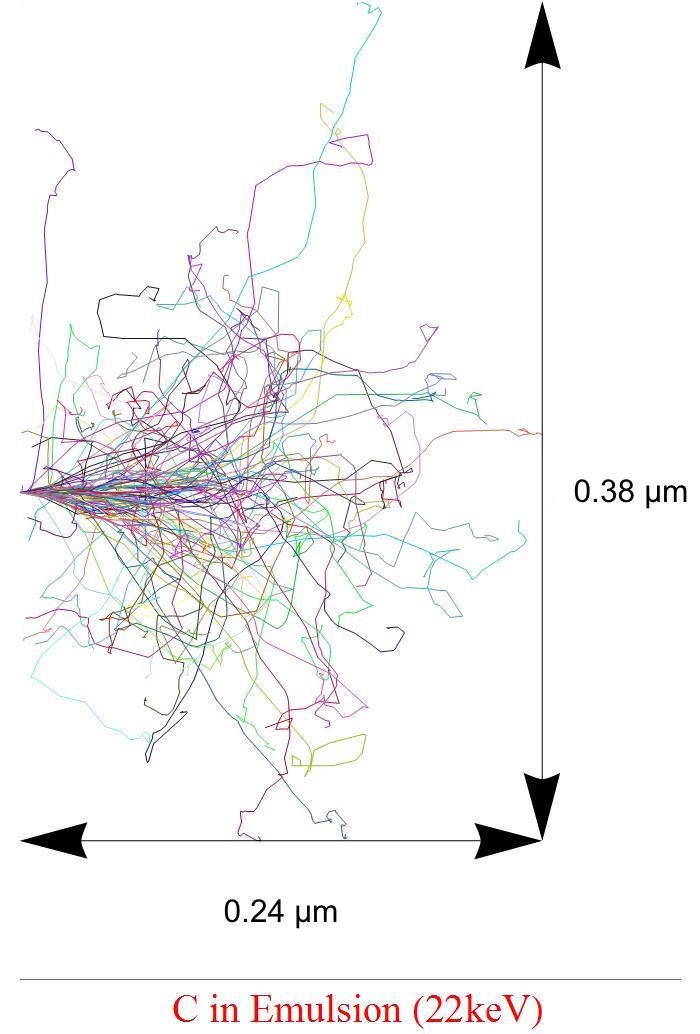}
			\includegraphics[width=0.328\linewidth]{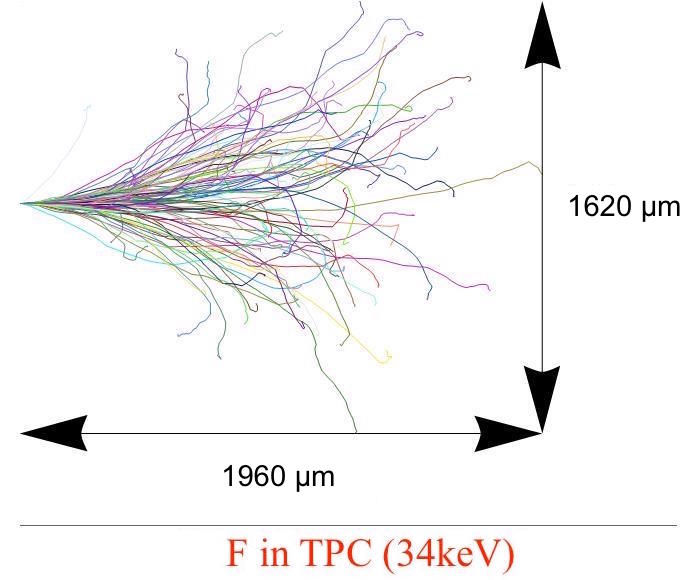}
	\caption{Development of the tracks of recoils with the maximum kinetic energy induced by a 1000\,GeV/c$^2$ WIMP. From left to right: $^{16}$O of 29\,keV in ZnWO$_4$, $^{12}$C of 22\,keV  in Emulsion, $^{19}$F of 34\,keV in the TPC gas mix. (SRIM simulations)}
	\label{fig:tracks}
\end{figure}

For each detector model, we simulated 7000 recoils induced by 1/10/100/1000\,GeV/c$^2$ WIMPs, leading to a 0.12/4.8/23/29\,keV maximum kinetic energy oxygen nucleus ($^{16}$O) in the
crystal detector model, to a 0.15/5/18/22\,keV carbon nucleus ($^{12}$C) in the emulsion
detector  and to a 0.10/4.6/26/34\,keV fluorine nucleus ($^{19}$F) in the considered TPC, using a WIMP velocity of 300\,km/s.
Figure \ref{fig:tracks} illustrates the development of the tracks of recoils induced by a 1000 GeV/c$^2$ WIMP, in the different media. For the emulsion and the crystal, the development of the tracks in the transverse direction (with respect to the initial direction of the recoil) can be larger than the development in the initial direction.

The range is defined as the distance traveled by the recoiling nucleus until it loses all its energy in the interaction medium (gas, emulsion or crystal), projected along the initial direction. 
Typically, for the three proposed strategies, the mean ranges of recoils induced by a 100\,GeV/c$^2$ WIMP are $0.04\pm 0.02$\,${\mu}$m  in Crystal, $0.09\pm 0.05$\,${\mu}$m  in emulsion, $800\pm 300$\,${\mu}$m  in the TPC gas mix. The range values for different WIMP masses are provided in Table \ref{tab:ranges}.

\begin{table}[htbp]

    \vspace{3mm}
	\centering
	\begin{tabular}{|c|c|c|c|}
		\hline
		& {$^{16}$O in Crystal} & {$^{12}$C in Emulsion} & $^{19}$F in TPC \\ \hline	
	    WIMP Mass (GeV/c$^2$) & \multicolumn{3}{c|}{Mean range $\pm$ std. deviation ($\mu$m)} \\ \hline
		1    & (1.78  $\pm$  0.92) 10$^{-3}$ & (4.15  $\pm$  1.93) 10$^{-3}$ & 13  $\pm$  6 \\ \hline
		10   & (8.64  $\pm$  4.49) 10$^{-3}$ & (3.25  $\pm$  1.73) 10$^{-2}$ & 170  $\pm$ 70 \\ \hline
		100  & (3.65  $\pm$  1.64) 10$^{-2}$ & (9.46  $\pm$  4.57) 10$^{-2}$ & 800  $\pm$ 300 \\ \hline
		1000 & (4.41  $\pm$  1.89) 10$^{-2}$ & (1.11  $\pm$  0.54) 10$^{-1}$ & 1040 $\pm$  360 \\ \hline
	\end{tabular}

	\caption{Mean range and standard deviation of the recoil tracks for the three different media, for WIMP masses from 1 to 1000\,GeV/c$^2$. \label{tab:ranges}}
\end{table}

\subsection{Angular distribution}\label{sub:angular-distribution}
\begin{figure}[!h]

    \vspace{10mm}
   	\centering
	\includegraphics[scale=0.42]{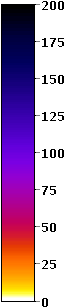}
	\includegraphics[scale=0.162,trim={115mm 10mm 115mm 9mm},clip]{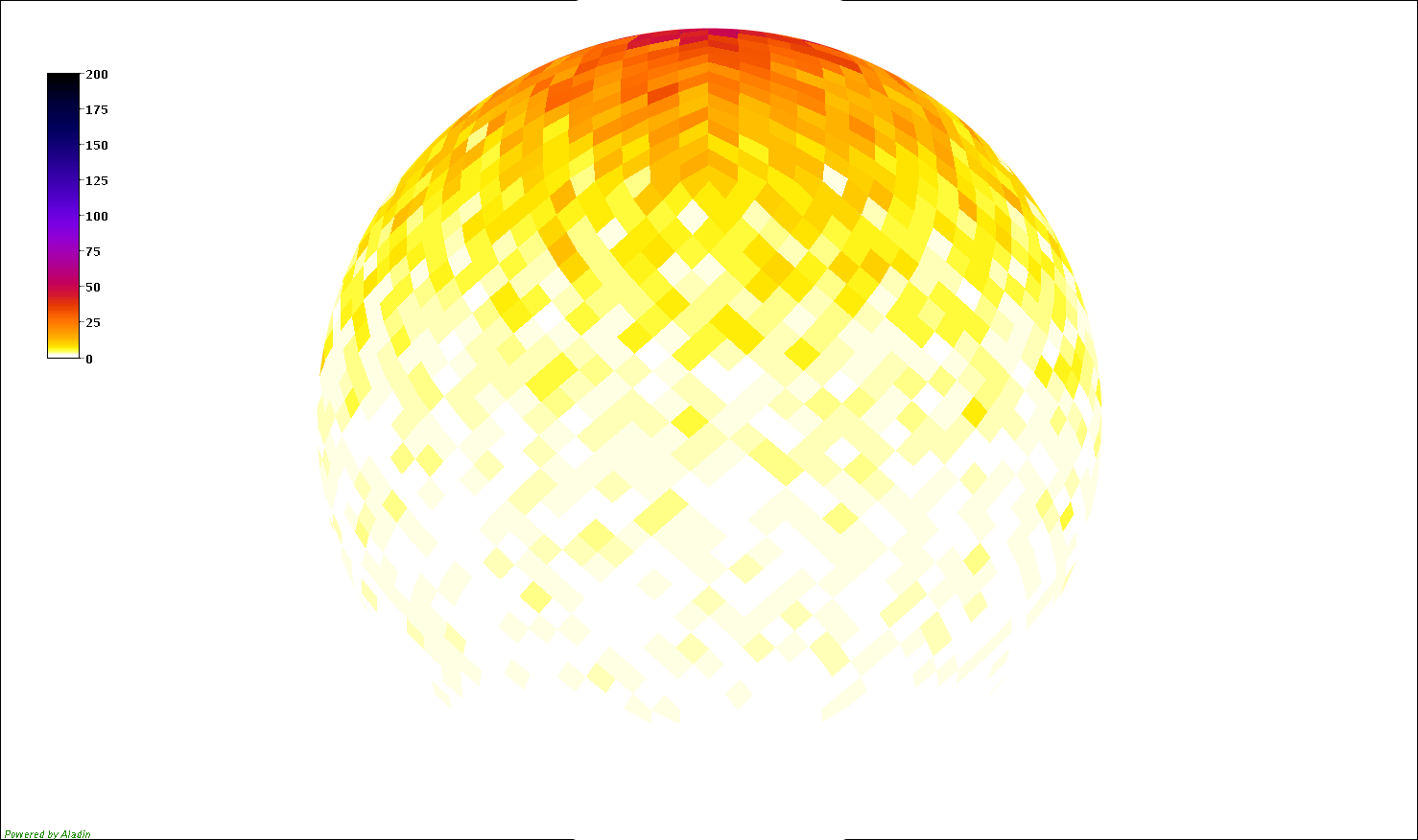}
	\includegraphics[scale=0.162,trim={115mm 10mm 115mm 9mm},clip]{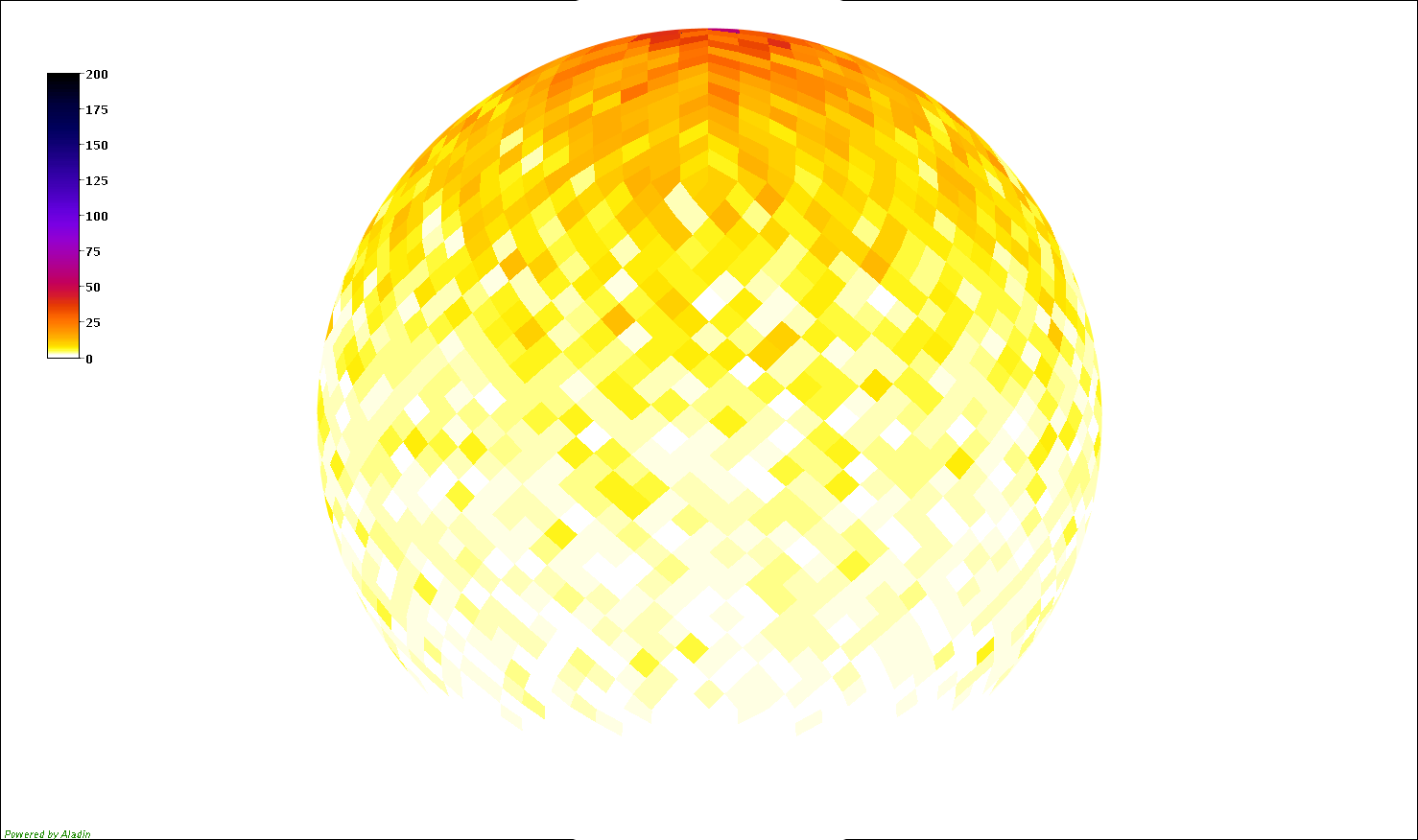}	
	\includegraphics[scale=0.162,trim={115mm 10mm 115mm 9mm},clip]{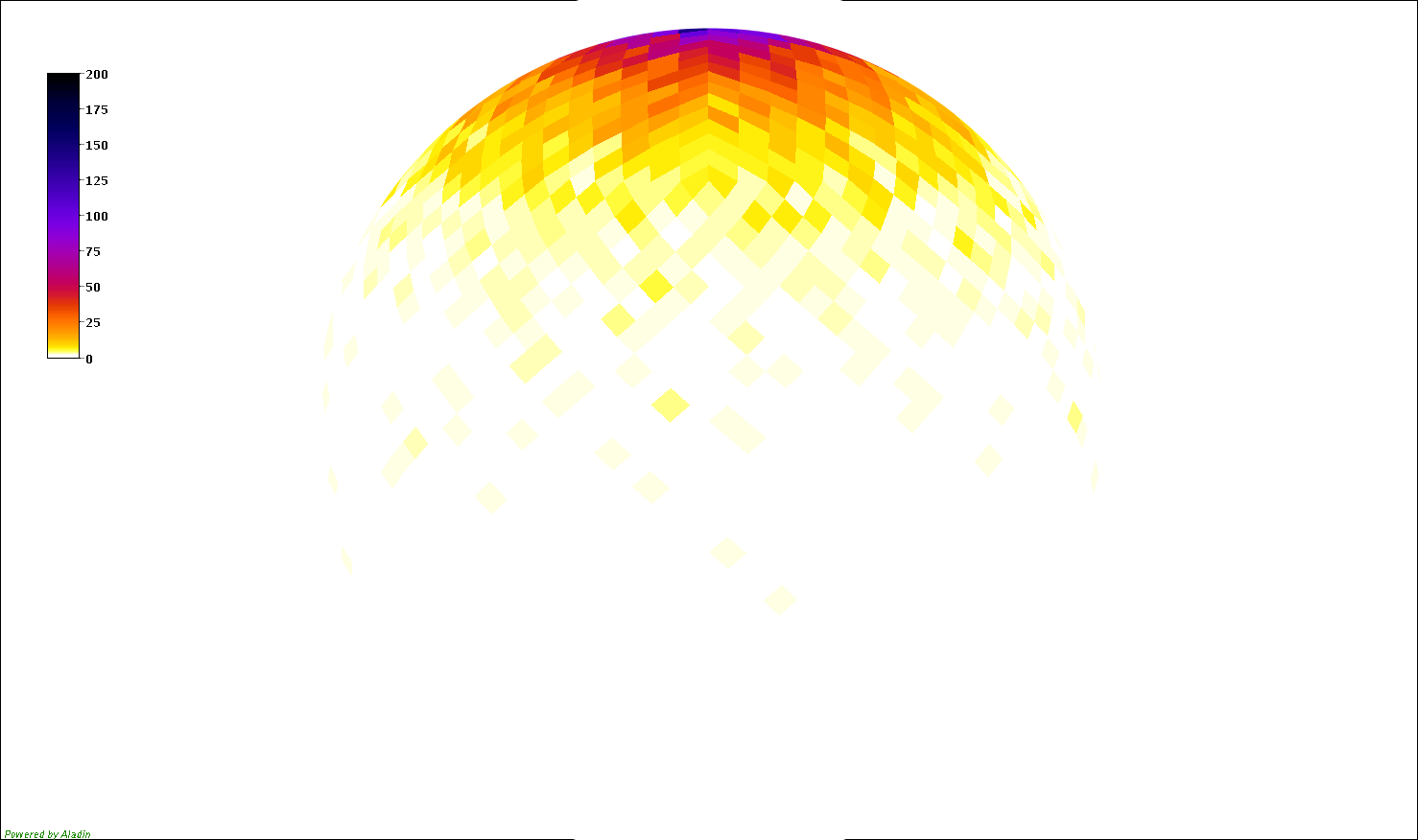}
	\includegraphics[scale=0.951,trim={9mm 0 0 0}]{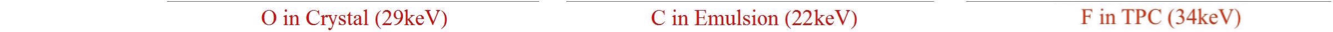}

    \vspace{3mm}
   	\caption{
		Projection on a sphere of the angles \{$\theta$, $\phi$\}  of 7000 simulated recoils tracks induced by a 1000 GeV/c$^2$ WIMP: crystal (left), emulsion (center) and TPC (right). The Z-axis, corresponding to the initial direction of the recoils, is oriented toward north. It can be noticed that for the TPC, the tracks remain in the upper hemisphere ($\theta<\pi/2$) while for emulsion and crystal, they extend beyond the equatorial plane ($\theta=\pi/2$).}
	
	\label{fig:angular_distribution_2D}
\end{figure}

\begin{figure}[!h]
	
	\centering
    \vspace{5mm}
	\includegraphics[width=0.495\textwidth]{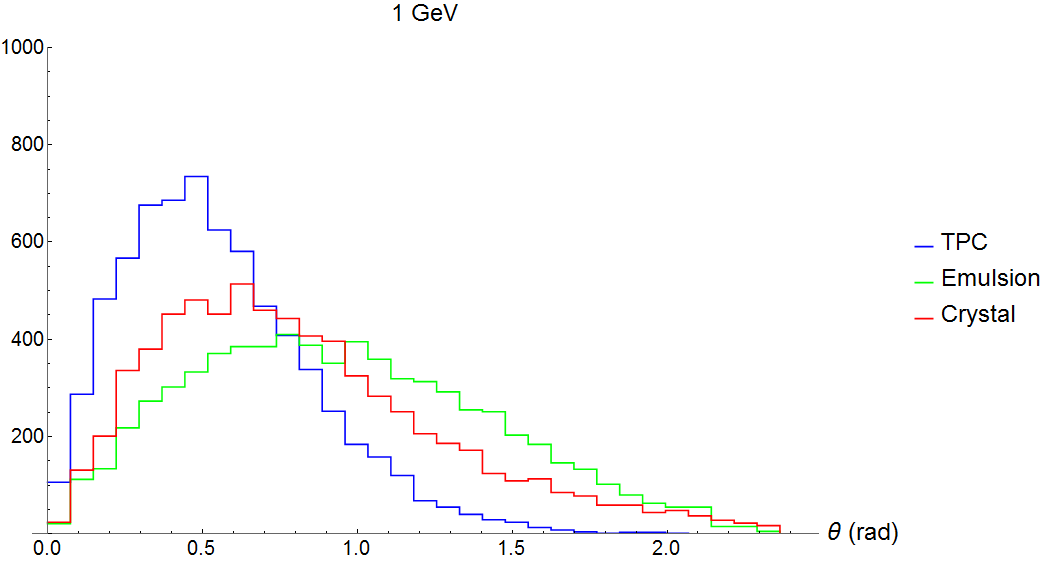}
	\includegraphics[width=0.495\textwidth]{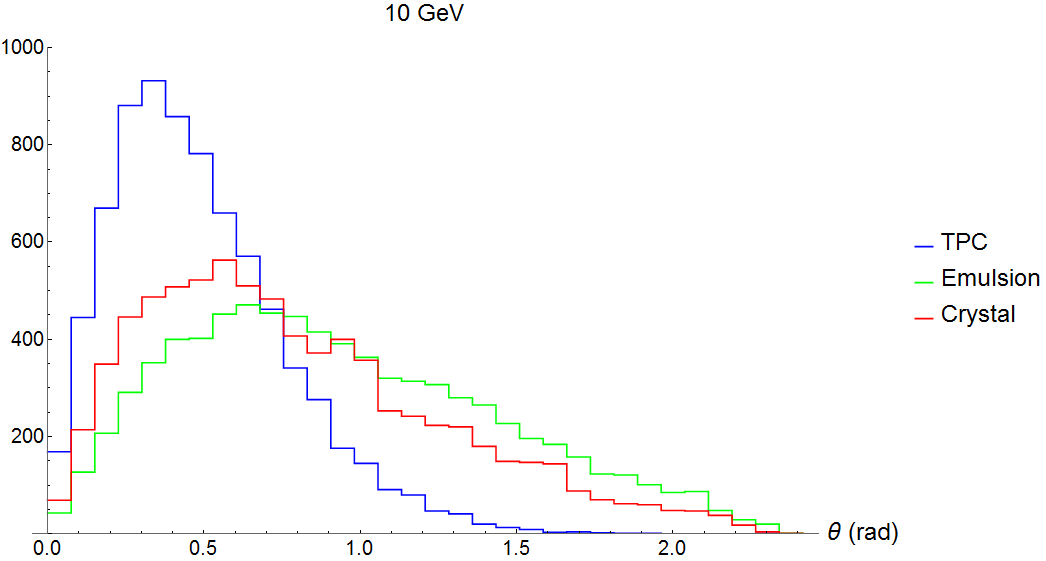}
	\includegraphics[width=0.495\textwidth]{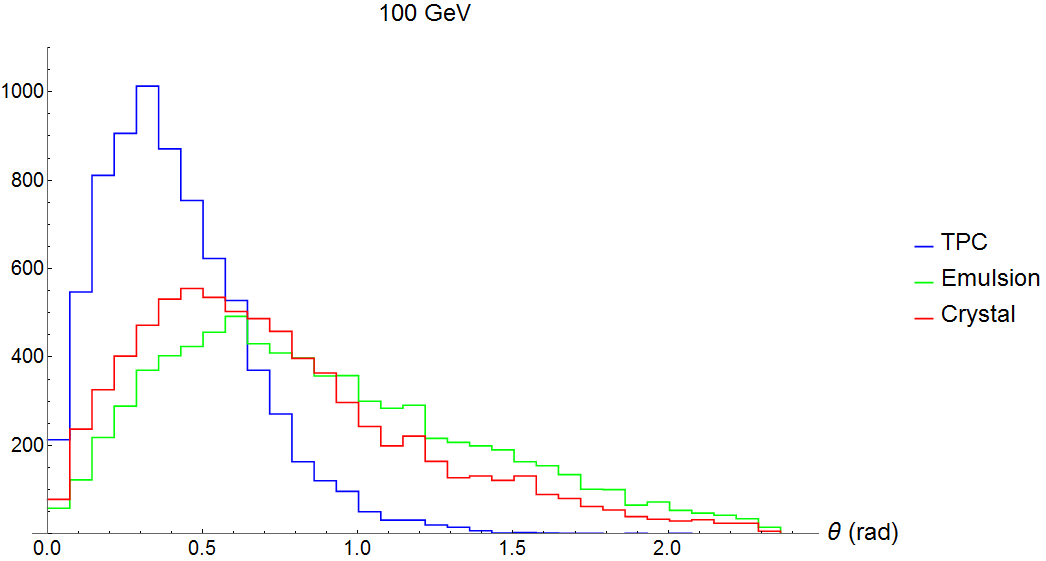}
	\includegraphics[width=0.495\textwidth]{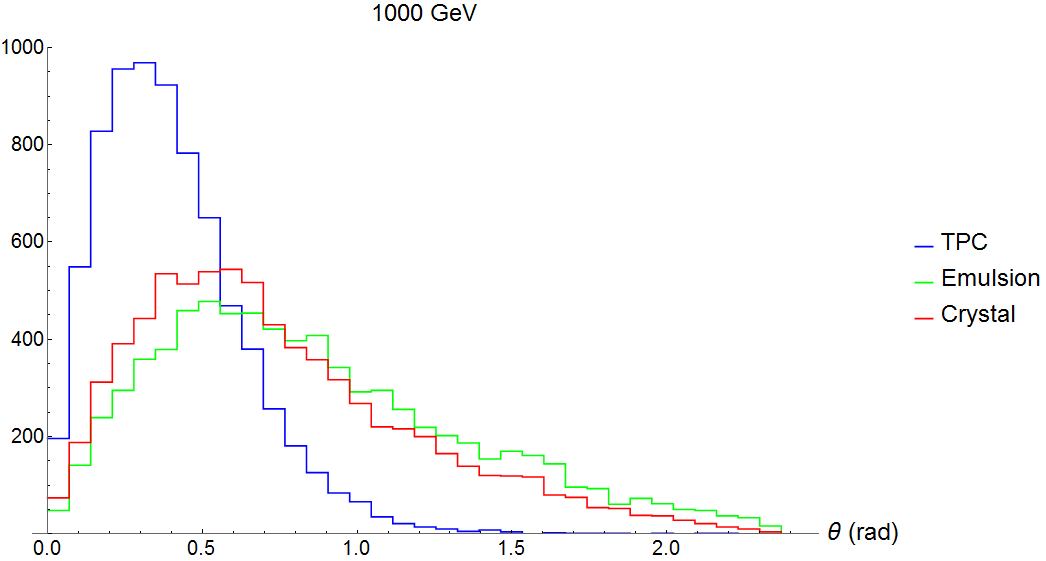}
    
    \vspace{3mm}
   	\caption{Comparison of the $\theta$ distributions for the three strategies -- gaseous TPC in blue, crystal in red, emulsion in green --  for 4 different WIMP masses: 1/10/100/1000\,GeV/c$^2$. 
	\label{fig:angular_distribution_1D} }
    \vspace{10mm}
\end{figure}

For each primary recoil track, we define the angle $\theta$ between the direction of the initial recoil and the line between the start point and the end of the track, and $\phi$ the angle around the initial direction ; $\theta$ and $\phi$ represent respectively the polar and azimuthal angles in spherical coordinates. 
Figure \ref{fig:angular_distribution_2D} shows the  projection of these angles on a sphere.
Figure \ref{fig:angular_distribution_1D} presents the distribution of $\theta$ for 7000 simulated recoils in crystal, emulsion and gas,  by 1 to 1000\,GeV/c$^2$ WIMPs.
In both Fig. \ref{fig:angular_distribution_2D} and Fig.  \ref{fig:angular_distribution_1D}, we can notice that the angular distributions for the TPC are less spread, and peak  at a lower $\theta$ value. The distributions for crystal and emulsion extend above $\theta = \pi/2$: the  corresponding tracks have lost the information of the initial direction.

\newpage

\subsection{A measure of the preservation of the directional information: observable \textit{D}}\label{sub:observable-d}

\begin{figure}[t]
\centering
\includegraphics[width=0.6\linewidth]{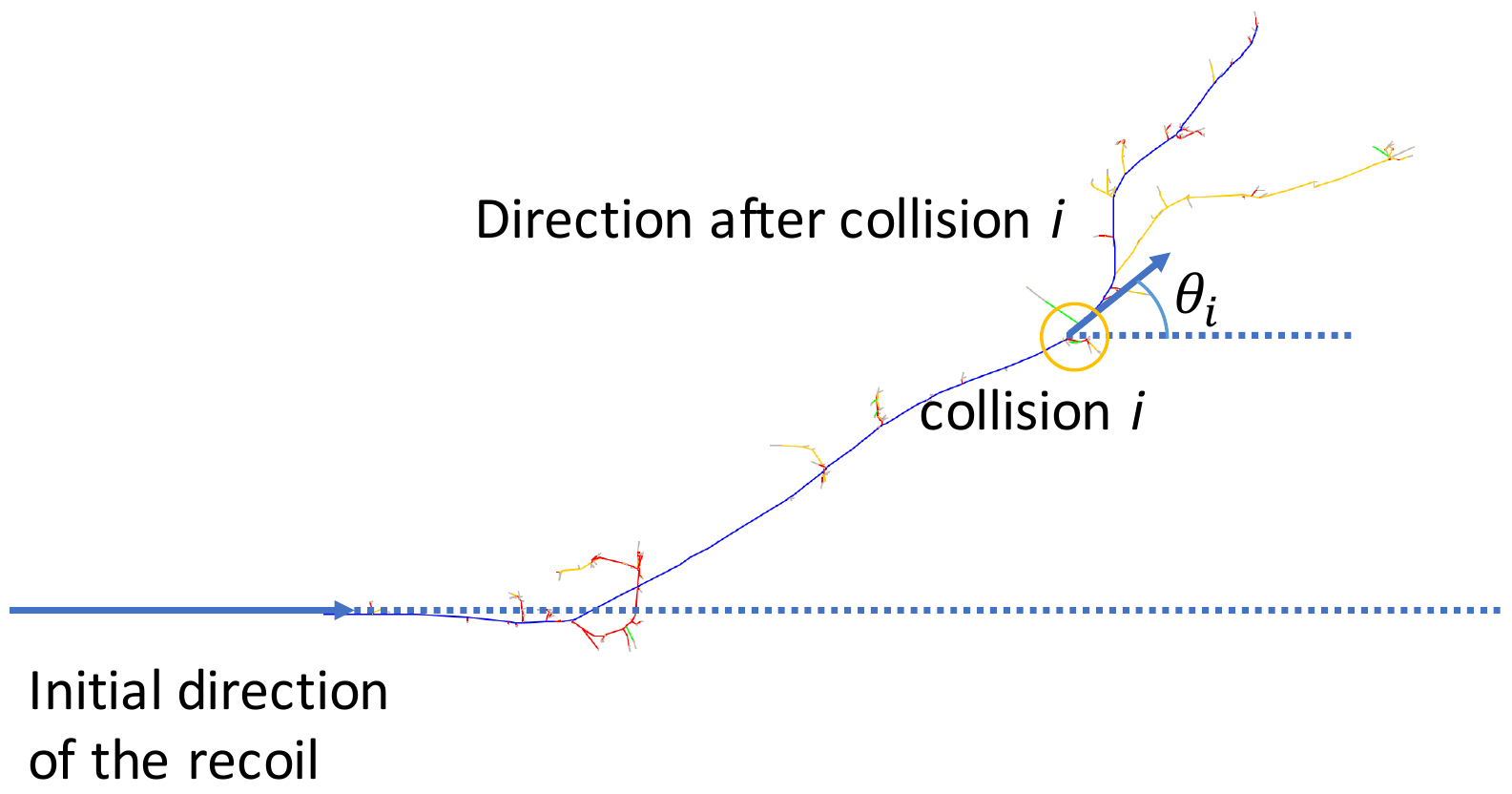}
\caption{A recoil track simulated with SRIM. The angle $\theta_i$ between the initial direction of the recoil and the direction after the $i$-th collision is shown.}
\label{fig:schematic_D}
\end{figure}

If we consider a simulated recoil track -- as is shown in figure \ref{fig:schematic_D} -- we quantify the preservation of the directional information using the figure of merit \textit{D}, defined as:

$$
    D = \frac{\sum_{i=0}^{N_{\rm collisions}} \cos(\theta_i)\cdot \Delta E_i}{\sum_{i=0}^{N_{\rm collisions}} \Delta E_i}  = \frac{\langle\cos(\theta) \cdot \Delta E\rangle_{\rm track}}{\langle\Delta E\rangle_{\rm track}}
$$

where: 
\begin{itemize}
	\item \(\theta_i\) is the angle between the direction of the initial recoil and the line joining two subsequent interaction points \(i\) and \(i+1\) (\textit{ie} the direction after the $i$-th collision); 
	\item \(\Delta E_i\) is the energy deposited by the recoil between \(i\) and \(i+1\);
	\item \(N_{\rm collisions}\) is the total number of
	interactions/collisions.
\end{itemize}

This figure of merit \textit{D} measures \emph{the mean direction of the track compared
to the primary direction of the recoil}. It takes values between $-1$ and 1: if close to 1, the ion is not much deviated and the initial direction is
preserved; if close to 0 (or negative), the initial direction is lost.
\mbox{(1 -- \textit{D})} would then quantify the loss of the directional information.

\begin{figure}[htbp]
	\centering
		\includegraphics[width=0.45\textwidth]{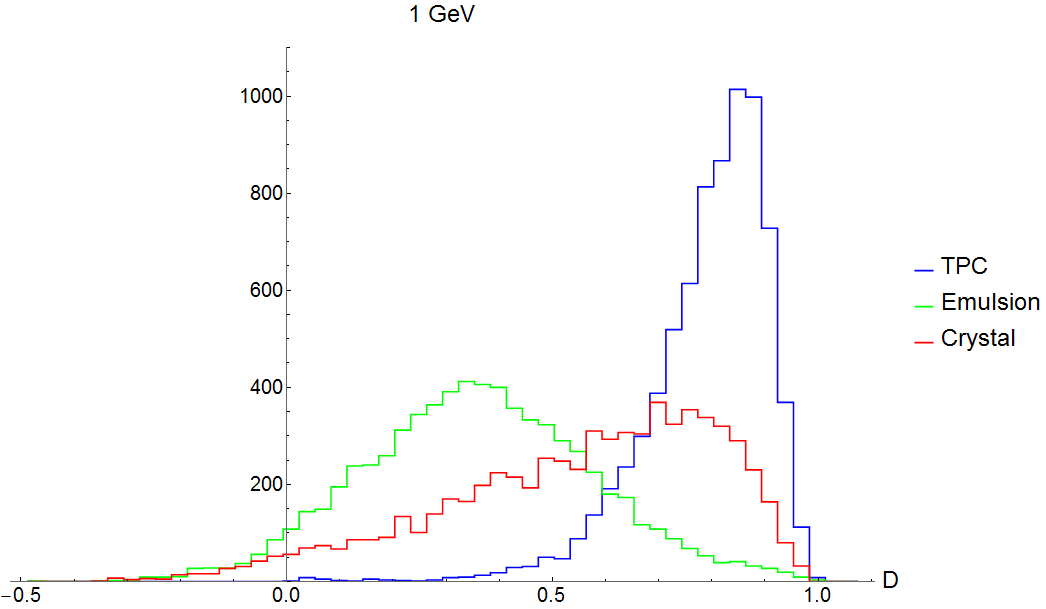}
		\includegraphics[width=0.45\textwidth]{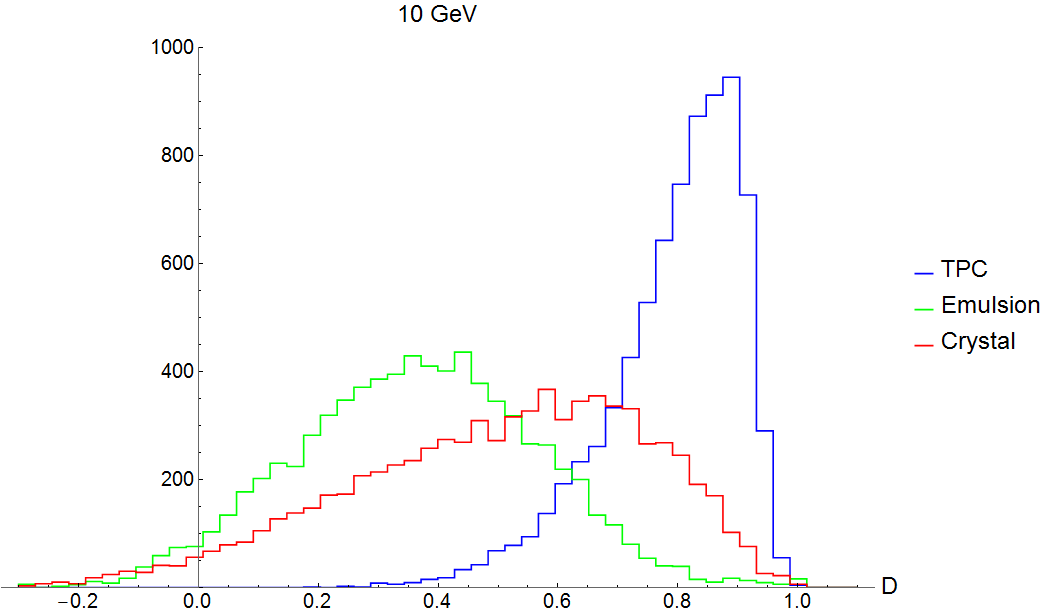}
		\includegraphics[width=0.45\textwidth]{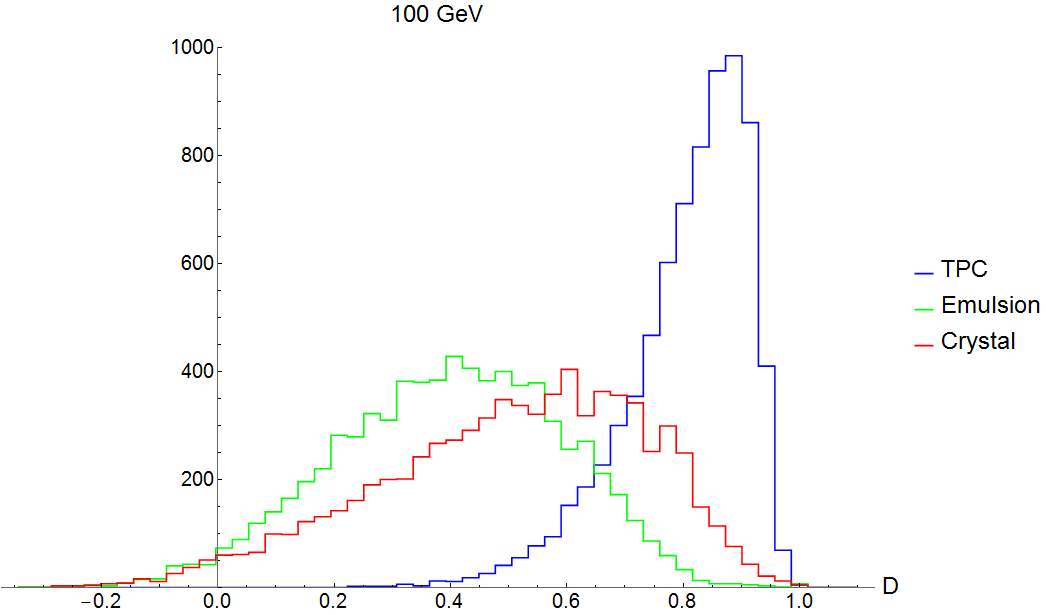}
		\includegraphics[width=0.45\textwidth]{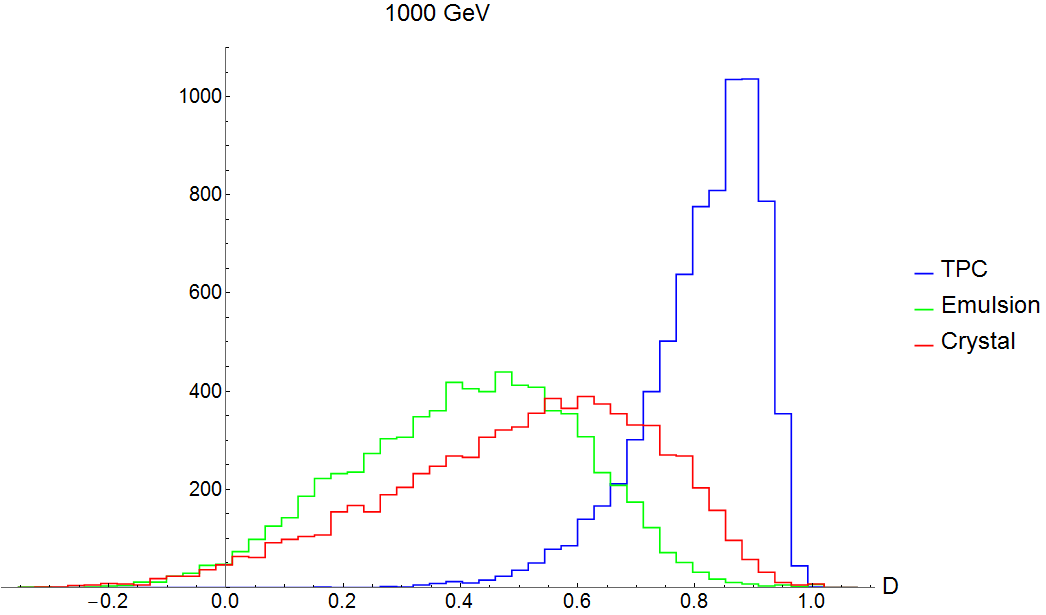}
		\vspace{2mm}
		
	\caption{Comparison of the directionality \textit{D} for the three strategies -- gaseous TPC in blue, crystal in red, emulsion in green --  for 4 different WIMP masses: 1/10/100/1000\,GeV/c$^2$. For all these WIMP masses, the simulations of the recoils in the different detector materials lead to higher \textit{D} values for a gaseous TPC (direction is better preserved); emulsion and crystal show similar, lower distributions of \textit{D}.
		\label{fig_D} }
\end{figure}

Figure \ref{fig_D} shows the distribution of the measure \textit{D} for nuclear
recoils induced by a \mbox{1/10/100/1000\,GeV/c$^2$} WIMP. For all the considered WIMP masses, the TPC presents a peaked distribution at high \textit{D} values: for instance, $\langle D\rangle = 0.9$ for a 100\,GeV/c$^2$ WIMP. The crystal and emulsion detectors show broader distributions, at lower values: $\langle D\rangle = 0.65$ for the crystal, 0.55 for the emulsion.  
This demonstrates that a low-pressure gas mix better preserves the initial direction of the recoil.

We noticed in the simulations that scattering on heavier nuclei leads to an earlier loss of kinetic energy of the primary recoiling nucleus. This suggests that for an optimal determination of directionality, having lighter components is crucial.

\newpage
\section{Discussion}\label{sec:discussion}
Using SRIM simulations, we compared how the recoil primary direction information is preserved in the detector materials for three different strategies for a Dark matter directional detection: anisotropic crystals, emulsion layers and low-pressure gaseous TPCs. 
Anisotropic crystals do not allow an event by event reconstruction of the tracks; yet showing how they preserve the direction information is useful to understand the prospective results of a directional measurement.
On the other hand, emulsion and TPCs detectors measure each recoil track; the ranges of the typical recoil tracks expected from the elastic scattering by a WIMP is consistent with the reconstruction resolution of their respective readouts.
We propose the use of a new observable, \textit{D}, to quantify the preservation of the initial direction information of the recoiling nucleus. This observable shows that among the three studied materials, a low pressure TPC gaseous mix is better suited for measuring the direction of WIMP-induced nuclear recoils.
In fact, dedicated measurements
with a calibration source producing ions with a known direction at a
given kinetic energy such as the COMIMAC facility \cite{muraz_comimac_2016} would allow to confirm the simulations presented
here.

\vspace{5mm}
We acknowledge F. Hosseini and Q. Riffard help during the phase of preliminary data analysis.
CC acknowdleges financial support from the Labex Enigmass.


\bibliography{directionality}

\bibliographystyle{JHEP}

\end{document}